\documentclass[twocolumn, deluxetables]{aastex63}
\usepackage{amsmath}
\usepackage[caption=false]{subfig}
\usepackage{multirow}
\usepackage{graphicx}
\graphicspath{{./}}

\shorttitle{Mrk 817}
\shortauthors{Zak et al.}

\begin{document}

\title{Fierce Feedback in an Obscured, Sub-Eddington State of the Seyfert 1.2 Markarian 817}

\author{Miranda K. Zak}
\email{mkzak@umich.edu}
\affiliation{Department of Astronomy, University of Michigan, 1085 South University Avenue, Ann Arbor, MI 48109, USA}

\author[0000-0003-2869-7682]{Jon M. Miller}
\affiliation{Department of Astronomy, University of Michigan, 1085 South University Avenue, Ann Arbor, MI 48109, USA}

\author[0000-0002-9356-1645]{Ehud Behar}
\affiliation{Department of Physics, Technion, Haifa 32000, Israel}

\author[0000-0002-0167-2453]{W. N. Brandt}
\affiliation{Department of Astronomy \& Astrophysics and the Institute for Gravitation and the Cosmos, The Pennsylvania State University, 525 Davey Lab, University Park,
PA, 16802, USA}

\author[0000-0003-2663-1954]{Laura Brenneman}
\affiliation{Center for Astrophysics, Harvard \& Smithsonian, 60 Garden Street, Cambridge, MA 02138, USA}

\author[0000-0002-2218-2306]{Paul~A.~Draghis}
\affiliation{Department of Astronomy, University of Michigan, 1085 South University Avenue, Ann Arbor, MI 48109, USA}

\author[0000-0002-0273-218X]{Elias Kammoun}
\affiliation{IRAP, Universite´ de Toulouse, CNRS, UPS, CNES, 9, Avenue du Colonel Roche, BP 44346, F-31028, Toulouse Cedex 4, France}
\affiliation{Dipartimento di Matematica e Fisica, Universit\`{a} Roma Tre, via della Vasca Navale 84, I-00146 Rome, Italy}

\author{Michael J. Koss}
\affiliation{Eureka Scientific, 2452 Delmer Street Suite 100, Oakland, CA, 94602-3017, USA }

\author[0000-0003-1621-9392]{Mark~T.~Reynolds}
\affiliation{Department of Astronomy, University of Michigan, 1085 South University Avenue, Ann Arbor, MI 48109, USA}
\affiliation{Department of Astronomy, Ohio State University,  140 W 18th Avenue, Columbus, OH, 43210, USA}

\author[0000-0002-0572-9613]{Abderahmen Zoghbi}
\affiliation{Department of Astronomy, University of Maryland, College Park, MD, 20742, USA}
\affiliation{HEASARC, Code 6601, NASA/GSFC, Greenbelt, MD, 20771, USA}
\affiliation{CRESST II, NASA Goddard Space Flight Center, Greenbelt, MD, 20771, USA}




\begin{abstract}
\noindent Markarian 817 is a bright and variable Seyfert-1.2 active galactic nucleus (AGN).  X-ray monitoring of Mrk 817 with the Neil Gehrels Swift Observatory in 2022 revealed that the source flux had declined to a lower level than recorded at any prior point in the then-19-year mission.  We present an analysis of deep XMM-Newton and NuSTAR observations obtained in this low flux state.  The spectra reveal a complex X-ray wind consisting of neutral and ionized absorption zones.  Three separate velocity components are detected as part of a structured ultra-fast outflow (UFO), with $v/c = 0.043^{+0.007}_{-0.003}$, $v/c = 0.079^{+0.003}_{-0.0008}$, and $v/c = 0.074^{+0.004}_{-0.005}$.  These projected velocities suggest that the wind likely arises at radii that are much smaller than the optical broad line region (BLR).  In order for each component of the outflow to contribute significant feedback, the volume filling factors must be greater than $f \simeq 0.009$, $f \simeq 0.003$, and $f \simeq 0.3$, respectively.  For plausible, data-driven volume filling factors, these limits are passed, and the total outflow likely delivers the fierce feedback required to reshape its host environment, despite a modest radiative Eddington fraction of $\lambda \simeq 0.008-0.016$ (this range reflects plausible masses).   UFOs are often detected at or above the Eddington limit; this result signals that black hole accretion has the potential to shape host galaxies even at modest Eddington fractions, and over a larger fraction of a typical AGN lifetime.  We discuss our findings in terms of models for disk winds and black hole feedback in this and other AGN.
\end{abstract}



\section{Introduction} \label{sec:intro}
Seyfert active galactic nuclei are powered by rapid accretion onto a massive black hole, roughly in an Eddington limit-scaled range of $\lambda_{Edd} = 0.001-0.1$ (where $\lambda_{Edd} = L_{bol}/L_{Edd}$, the bolometric luminosity divided by the Eddington luminosity).  Seyfert-1 AGN are relatively ``unobscured'' -- emission from the central engine is seen directly and through reprocessing in the optical ``broad line region'' or BLR.  In contrast, Seyfert-2 AGN are obscured by an extended ``torus'' region, so that the central engine and broad optical lines are not seen directly (see, e.g., Antonucci 1993; for a helpful discussion of geometries, also see Peterson et al.\ 2004).  Although a classic picture of the torus envisioned a parsec-scale geometry, recent dust reverberation mapping finds that its inner edge is likely only a few times larger than the optical BLR (e.g., Minezaki et al.\ 2019, Yang et al.\ 2020).

Markarian 817 is a nearby, $z=0.03145$ Seyfert-1.2 AGN (Strauss \& Huchra 1988; Koss et al.\ 2017; also see Miller et al.\ 2021).  When it was studied within the CfA Seyfert Sample (Huchra \& Burg 1992), its X-ray flux was a factor of 40 lower than when it was observed with XMM-Newton in 2009, though its UV flux was broadly comparable (Winter et al.\ 2011).  This behavior is unusual among Seyfert-1 AGN, though it is qualitatively consistent with some aspects of the transient ``obscurers'' seen in, e.g., NGC 5548 and NGC 3783 (Kaastra et al.\ 2014, Mehdipour et al.\ 2017).  

The mass of the black hole in Mrk 817, $M = 8.12^{+0.73}_{-0.67} \times 10^{7}~M_{\odot}$, has recently been refined through continued optical BLR reverberation mapping (Lu et al.\ 2021).  Recent work by Kara et al.\ (2021) and Homayouni et al.\ (2023) use a smaller black hole mass of $M \sim 3.85\times 10^{7}~M_{\odot}$ based on some similarities to NGC 5548.  The fact that optical reverberation mapping sometimes detects a clear lag in Mrk 817 demands that reprocessing of the centrally emitted continuum by the BLR happens during some intervals or states.  However, a long and intensive X-ray and UV continuum monitoring campaign with the Neil Gehrels Swift Observatory found no link between X-ray and UV emission (Morales et al.\ 2019), indicating that reprocessing might frequently be disrupted.  A deep XMM-Newton observation in 2021 detected relativistic reflection from the inner disk and strong, ionized absorption; the former indicates that the inner disk is seen clearly, while the latter is consistent with a disk wind that could disrupt reprocessing in the optical BLR (Miller et al.\ 2021).  This interpretation was confirmed in follow-up observations that included UV spectroscopy (Kara et al.\ 2021).  

The obscuration that was revealed with XMM-Newton was found to have a relatively high column density ($N_{\rm H} = 4.0^{+0.7}_{-0.5}\times 10^{22}~{\rm cm}^{-2}$), a large covering factor ($f_{cov} = 0.93\pm 0.02$), a modest ionization ($log \xi = 0.5^{+0.5}_{-0.4}$), and to potentially be very fast ($v/c = -0.12^{-0.02}_{+0.09}$ (Miller et al.\ 2021).  Fast winds are particularly important for understanding the interplay of the corona, inner accretion disk, and BLR in AGN.  They are of more cosmological importance because such flows may contribute the feedback required to clear the host galaxy bulge of gas for later star formation.   Simulations indicate that the kinetic power in winds needs to represent approximately 5\% of the radiative luminosity to achieve such impacts (e.g., Di Matteo et al.\ 2005).  However, it remains an open question whether such flows are only possible in rare super-Eddington phases, or if they might be generated more regularly at modest Eddington fractions.   X-ray spectroscopy is the tool that is likely best suited to revealing extreme winds (e.g., Gallo, Miller, \& Costantini 2023).

Cold and moderately ionized gas absorbs and scatters soft X-rays (e.g., $E \leq$2-3~keV) more efficently than hard X-rays (e.g., $\geq$4--10~keV).  Episodes of strong X-ray obscuration in AGN can therefore be identified through the use of X-ray hardness ratios.  Momentary changes can be due to partial or full eclipses (see, e.g., Gallo et al.\ 2021), but longer changes can indicate sustained wind obscuration.  Figure 4 in Miller et al.\ (2021) shows that wind obscuration episodes are indeed marked by rises in X-ray hardness in Swift/XRT monitoring of Mrk 817.  Continued Swift monitoring of Mrk 817 in 2022 revealed that the source was an order of magnitude fainter than even in the intensive campaign in 2017--2018 (Zak et al.\ 2021).  On this basis, we obtained additional observations with NuSTAR, XMM-Newton, and Swift (Miller et al.\ 2022), finding that the ionized obscuration had increased by an order of magnitude over any prior measurements.  

Herein, we report on detailed fits to the X-ray spectra from NuSTAR and XMM-Newton.  The data are particularly complex and require a series of absorbers spanning a broad range of ionization, column density, and outflow speed.  Importantly, the data strongly prefer very high velocities, and broadly suggest that the fastest wind components originate closest to the black hole.  Section 2 details the observations, and our reduction of the data.  In Section 3, we present our analysis and results.  A discussion of our findings, the strengths and weaknesses of our methods, and future directions is presented in Section 4.  

\section{Observations and Data Reduction} \label{sec:obs}

\subsection{NuSTAR}
NuSTAR observation 90801608002 started on 07 April 2022 at 20:26:09 (UT) and achieved a total exposure of 46.9~ks.  Using NUSTARDAS v2.1.1 and CALBD v202204026, we ran the NuSTAR data analysis pipeline ``nupipeline.''  We applied standard screening for passages through the South Atlantic Anomaly (SAA).  Source spectra, background spectra, and responses from each of the two focal plane modules (FPMA, FPMB) were generated using the tool ``nuproducts.''  The source region and background region were circular and had radii of 75 arc seconds and 110 arc seconds, respectively. There is no contribution from nearby sources. Background regions were extracted from the same detector quadrant as the source. The spectrum from FPMA had 4475 counts, with 732 background counts, while the FPMB spectrum had 4414 counts and 1107 background counts.  The spectra were rebinned using ``ftgrouppha'' with the Kaastra \& Bleeker (2016) optimal binning algorithm within the FTOOLs suite (Blackburn et al.\ 1999, see http:\/\/heasarc.gsfc.nasa.gov).

\subsection{XMM-Newton}
XMM-Newton observation 0911990201 started on 20  April 2022 at 22:46:18	(UT) and yielded a total exposure of 128.8~ks.  The data were reduced using the tools and packages within SAS version 1.3 (xmmsas\_20211130\_0941\_20.0.0; see Gabriel et al.\ 2004), and corresponding calibration files.  The EPIC cameras were operated in ``Large Window'' mode with the ``medium'' optical blocking filter.  

We produced a calibrated event list for the EPIC-pn camera by running ``epproc.''  Selecting a region on the same chip as Mrk~817, we created a background pn light curve to identify periods of soft proton flaring, and eliminated these periods by creating and applying an additional good time interval (GTI) file via ``tabgtigen.''  After further standard filtering (``FLAG=0'' and ''PATTERN$\leq$4''), we created pn source and background spectra and responses (via ``rmfgen'' and ''arfgen'').  Source and background regions were 43.5'' in radius.  The net pn exposure time was 98.1~ks. The pn spectrum had 61883 counts in the 0.3-10 keV range, and the associated background had 1773 counts.

Similarly, we created calibrated event lists for the MOS-1 and MOS-2 cameras by running ``emproc."  Background regions were again selected to create light curves that were used to identify and then excise instrumental flaring intervals via ``tabgtigen.''  Applying standard filtering (``FLAG=0'' and ''PATTERN$\leq$12''), source and background regions were created for each MOS camera using circular regions with a radius of 43.5''.  The net MOS-1 and MOS-2 exposure times were both 113.3~ks. The MOS-1 spectrum had 15174 counts from 0.3-10 keV, and there were 736 background counts. MOS-2 had 16741 counts in the same energy range, with 732 background counts.

The spectra from the RGS units appear to broadly confirm the EPIC-pn results detailed below, but offer no improvements to the constraints.  Data from the OM camera were obtained in a special readout mode to search for rapid variations in the UV light curve; these results will be reported in a separate paper.

\subsection{Swift}
Swift/XRT observations of Mrk~817 obtained between May 2007 and June 2022 were reduced to create light curves of the 0.3--10.0~keV count rate and 2.0--10.0~keV/0.3--2.0~keV hardness ratio.  A large gap in the monitoring of Mrk 817 occurred for much of 2019 and 2020, and the majority of observations occurred in 2017--2018 and 2021--2022. The total exposure time included in the light curves is 566~ks; individual observations were typically 0.9~ks. 

\section{Analysis and Results}

\subsection{Swift Light Curves}

Figure 1 shows the Swift/XRT count rate observed from Mrk~817 versus MJD, and the 2.0--10.0~keV/0.3--2.0~keV X-ray hardness ratio versus MJD.  The combination of these curves can serve as a coarse indicator of genuine flaring episodes versus obscuration events.
Between MJD 57700--58400, the count rate is highly variable on all time scales, and the hardness ratio for this period is typically very low.  This is consistent with genuine flaring driven by changes in the accretion rate (and not, e.g., ``uncovering'' events that allow the soft flux to emerge in rare intervals).  However, there are brief periods when the total count rate dips to 0.2~counts~s$^{-1}$ {\it and} the hardness ratio exceeds 0.5; these 
likely represent wind-driven obscuration events.  After MJD 59200, the baseline count rate observed from Mrk~817 is very low, flares are comparatively muted and diminish in extremity over time, and the hardness ratio is both very high and variable.  This likely indicates that the source entered a state with high but variable obscuration, consistent with a strong wind.  It is notable that the XMM-Newton and NuSTAR observations that we obtained occurred after an interval with an even higher hardness ratio close to MJD 59625, potentially indicating that the obscuration was even stronger at that time.

\subsection{XMM-Newton}

The source light curves from the three EPIC cameras on the 0.3--10.0~keV band are shown in Figure 2.  Instances of zero flux are due to excision of instrumental flaring periods; note that some apparent dips are not coincident between the cameras.  The count rate in each camera is remarkably steady; moreover, the count rate in various soft hand hard bands is also steady.  We therefore elected to fit the time-averaged spectrum from each.  Prior to spectral fitting, each spectrum was binned to have a signal-to-noise ratio of 5.0 using the tool ``ftgrouppha.''  All spectral fitting was performed using XSPEC version 12.11.1 (Arnaud 1996).

The left panel in Figure 3 shows a simple comparison of the EPIC-pn spectrum obtained in 2009 and (first analyzed by Winter et al.\ 2011), and the spectrum that we obtained in 2022.  Whereas the 2009 spectrum shows no clear signs of ionized absorption, and only weak evidence of even a narrow Fe~K emission line, the 2022 spectrum is extremely complex.  The new spectrum has more continuum emission than is typically observed in heavily obscured Seyfert-2 AGN, suggesting that the absorption is patchy, ionized, or both.  Several features within the new spectrum suggest that strong emission lines contribute to the total observed flux.  

We first attempted to jointly fit the new EPIC spectra with a simple model, consisting of neutral partial covering of a cut-off power-law direct continuum component, and linked distant neutral reflection (in XSPEC, $const\times tbpcf\times cutoffpl + pexmon$).  The parameters of the $tbpcf$  component include the redshift of the source, the equivalent neutral hydrogen column density (${\rm N}_{\rm H}$), and the dimensionless covering factor ($0\leq f\leq 1$).  The key parameters of the $pexmon$ component (see Nandra et al.\ 2007) include the power-law photon index (linked to the same parameter in the cut-off power-law component), the cut-off energy (linked to that in the cut-off power-law component, and fixed at ${\rm E}_{cut} = 100$~keV based on Fabian et al.\ 2015), the reflection fraction ($R = \Omega/2\pi$, and set to negative values to only include reflected emission), the inclination of the reflector (arbitarily fixed to $\theta = 30^{\circ}$), and the model flux normalization (linked to that of the cut-off power-law component).  We left the iron abundance fixed at the default (solar) value.   The multiplicative constant acted as a flux normalization factor between the EPIC cameras, with a value of 1.0 fixed for the pn and variable values for the MOS-1 and MOS-2 cameras.

This simple model is also unacceptable; see the righthand panel in Figure 3.   Strong absorption and emission lines remain unmodeled across the full pass band.  This is confirmed by the fit statistic: $\chi^{2} = 3151.1$ for $\nu = 1760$ degrees of freedom.  The model succeeds in describing some of the complexity in the Fe K region, suggesting that neutral (or, low-ionization) partial covering likely plays a role.  However, even the basic continuum is implausible: at $\Gamma = 2.71$, the implied direct power-law would be far steeper than a typical value of $\Gamma = 1.8$ (e.g., Nandra et al.\ 2007).  This likely indicates that the continuum is trying to account for low-energy flux that may actually be a pseudocontinuum of emission lines.  (The normalizing constants for the MOS-1 and MOS-2 were 0.91 and 1.07, respectively.)
 
Motivated by the fact that Seyfert-2 obscuration is dominated by cold, largely neutral gas, we next added an additional neutral partial covering component to this model (in XSPEC, $const\times tbpcf\times tbpcf\times cutoffpl + pexmon$).  This change yielded no improvement in the fit statistic, and the total column in the two neutral absorbers was equivalent to that in the single absorber in the prior, sipmler model.  This signals that the obscuration cannot be described exclusively in terms of cold absorption, even allowing for different partial covering factors.

We next considered a model that added an ionized partial covering absorber, using the model \texttt{zxipcf} (in XSPEC, $tbpcf\times zxipcf\times cutoffpl + pexmon$).  The parameters of this component include a velocity shift, a covering factor, the ionization of the absorber (${\rm log} \xi$, where $\xi = L/nr^{2}$, and $L$ is the bolometric luminosity, $n$ is the gas number density, and $r$ is the distance to the ionizing source).  Fits with the ionized absorber fixed at the redshift of the host yield a significant improvement ($\chi^{2} = 3023.0$ for $\nu = 1757$), but many of the same absorption and emission features remain, and the power-law index is unchanged.  Moreover, the ionization level is extremely low, ${\rm log}\xi = -0.24$, and the column density of the cold neutral gas and ionized gas (${\rm N}_{\rm H} = 4.4\times 10^{23}~{\rm cm}^{-2})$ roughly sum to that in the fit with just a single neutral absorber.  Nevertheless, the improvement suggests that ionized absorption is present.

While flexible, the ``zxipcf'' component has an important limitation: it samples nine decades in ionization with only 12 steps (e.g., Reynolds et al.\ 2012), so it can only give a coarse idea of the gas properties.  It also lacks a corresponding emission component, which is unphysical since the same gas that absorbs flux {\em must} re-emit it isotropically.  Wind re-emission of this kind is seen clearly in stellar-mass black holes (Miller et al.\ 2015, 2016; Trueba et al.\ 2019).

For these reasons, we created a grid of photoionized absorption and re-emission models using XSTAR (see, e.g., Kallman \& McCray 1982, Bautista \& Kallman 2001).  As an input spectrum, we assumed a bolometric source luminosity of $L = 1\times 10^{44}~{\rm erg}~{\rm s}^{-1}$, divided between a $T = 25,000$~K blackbody and a $\Gamma=1.7$ power-law in a 10:1 ratio to mimic a typical AGN spectral energy distribution (the power-law was artificially diminished to zero at low energy).   We also assumed solar abundances and a turbulent velocity of $v_{\rm turb} = 300~{\rm km}~{\rm s}^{-1}$.  The resultant XSPEC ``table'' models consist of 80 grid points spanning  $-2 \leq {\rm log}~\xi\leq 6$ and 40 grid points spanning $1.0\times 10^{20}~{\rm cm}^{-1} \leq {\rm N}_{\rm H}\leq 6.0\times 10^{23}~{\rm cm}^{-2}$.  

We next added pairs of ionized XSTAR absorption and re-emission components to the model.  The column density and ionization were tied within each pair, the absorption was allowed to be blue-shifted relative to the host galaxy, and the emission was fixed at zero with respect to the host galaxy.  This broadly corresponds to a situation wherein wind zones are launched in symmetric annuli, causing velocities to largely cancel in emission though a blue-shift is seen in absorption.  Velocities seen in emission are less likely to cancel if a source is viewed at high inclination, but prior reflection modeling in Mrk 817 suggests a low inclination (Miller et al.\ 2021).  At high spectral resolution, velocity broadening may be evident in the re-emission components.   Additional zones of paired absorption and emission were added until there was no significant improvement in the $\chi^{2}$ statistic, or until the parameters of a pairing could not exclude the minimum column density (in the case of absorption) or normalization (in the case of emission).   For completeness, we also included neutral absorption from the Milky Way via a $tbabs$ component, though its effect is negligible (a value of $N_{H} = 1.06\times 10^{20}~{\rm cm}^{-2}$ was adopted as per the HI4PI collaboration measurement; Bekhti et al.\ 2016).

Our final, best-fit model consists of neutral partial-covering absorption, three pairs of coupled photoionized absorption and re-emission, one unpaired photoionized emission component, a cut-off power-law direct continuum, and pexmon.  In XSPEC terms, this model can be written as: $tbabs\times (tbpcf\times abs_{1}\times abs_{2}\times abs_{3}\times cutoffpl + pexmon + emis_{1} + emis_{2} + emis_{3} + emis_{4})$.  The unpaired emission component has a very low ionization and likely corresponds to extended, diffuse emission that has been tied to the optical narrow-line region (NLR) in Chandra imaging studies of more proximal Seyferts (e.g., NGC 4151, Wang et al.\ 2011).  The details of this model are given in Table 1; the fit and model are shown in Figures 4 and 5.  A fit statistic of $\chi^{2} = 2089.39$ is measured for $\nu = 1746$ degrees of freedom. Even this model is not formally acceptable, likely owing to remaining wind complexities.  

The results of fits with this model clearly indicate that the central engine of Mrk~817 was obscured by a very complex disk wind.  Both the neutral and ionized absorption is optically thin (within our line of sight, at least).  However, each component has a major impact on the spectrum, with columns in excess of $N_{\rm H} = 10^{23}~{\rm cm}^{-2}$.  Two of the absorbers can be described as highly ionized, with ${\rm log}\xi = 3.31^{+0.04}_{-0.01}$ and ${\rm log}\xi = 3.66\pm 0.02$, but the third is very highly ionized: ${\rm log}\xi = 5.5^{+0.2}_{-0.1}$.  Its upper limit is consistent with that of the table grid.  While this component is primarily important in the Fe~K band, it contributes to absorption and emission across the pass band and cannot be neglected.

Most importantly, perhaps, the ionized absorption components have extremely high blueshifts, between $v/c =$0.036--0.0798 (including errors).  These components therefore rank as ``ultra-fast'' outflows (see, e.g., Tombesi et al.\ 2013).  The highly shifted Fe~K absorption line at 7.25~keV in Figures 4 and 5 is likely the clearest single signature of this UFO; other key lines are blended at lower energy.   The same absorption shapes the low energy spectrum and the velocities are also determined by those data; that part of the spectrum is just more complex.  The key details of the wind components are treated separately in the following section.

The best-fit continuum gives a total obscured and reprocessed flux of $F_{obs} = 2.1\pm 0.5\times 10^{-12}~{\rm erg}~{\rm cm}^{-2}~{\rm s}^{-1}$ in the 0.3--10.0 keV band.  When the complex absorption is removed, the unobscured direct power-law flux is $F_{unobs} = 8.0\pm 0.5\times 10^{-12}~{\rm erg}~{\rm cm}^{-2}~{\rm s}^{-1}$.  This is nearly four times larger than the obscured flux that is actually received at the detector, and more comparable to the flux levels previously observed in this source (e.g. Morales et al.\ 2019, Miller et al.\ 2021).  This suggests that the drop in the Swift/XRT count rate seen in Figure 1 is driven by a major change in obscuration, and that changes in the mass accretion rate likely play a secondary role.

\subsection{NuSTAR}
To enable direct comparisons, we fit the same model described in the previous section to the FPMA and FPMB spectra from NuSTAR observation 90801608002. The two spectra were fit jointly from 3-25 keV with a constant allowed to float between them.  Redshifts, ionization parameters, and photoionized emission component normalizations were frozen owing to the limited resolution and pass band of the NuSTAR detectors. The column density of zone 3 was also frozen due to its unpredictability from fit to fit. The neutral gas partial covering factor, column densities (excluding the one from the highly ionized absorption/emission pair), photon index, reflection fraction, and constant were free to change within the fit.  

The resulting fit statistic was $\chi^{2} = 190.88$ for $\nu = 133$ degrees of freedom. The confidence interval for each of the free parameters contained the value of the corresponding value from XMM-Newton (see Table 1). In the NuSTAR fit, the photon index is $\Gamma = 1.8^{+0.4}_{-0.2}$; this is formally consistent with the value measured in fits to the XMM-Newton/EPIC spectra, indicating that the proper continuum has been measured despite the complexity of those data.

The partial covering column density measured in the NuSTAR spectra is ${\rm N}_{\rm H} = 30^{+40}_{-20}\times 10^{22} \rm ~ cm^{-2}$, and the covering fraction is $0.8^{+0.1}_{-0.2}$. The column densities of the absorption components in zones 1 and 2 are ${\rm N}_{\rm H} = 10^{+20}_{-7} \times 10^{22}$ and ${\rm N}_{\rm H} = 10^{+10}_{-9} \times 10^{22}\rm ~ cm^{-2}$, respectively. The first is not within the errors of the XMM-Newton fit, but the best-fit XMM-Newton value is contained within its uncertainty, and it is within the realm of reasonable values, especially considering the  often rapid and dramatic variation in photon count rate and hardness ratio illustrated in Figure 1. The power-law continuum normalization is $3^{+3}_{-1} \times 10^{-3}~{\rm ph}~{\rm cm}^{-2}~{\rm s}^{-1}~{\rm keV}$ (at 1 keV), and the reflection fraction is $R = -0.4^{-0.3}_{+0.3}$. The floating constant is $0.96 \pm{0.04}$. A plot of the spectra and the fits can be found in Figure 7.

\subsection{Broad Spectral Energy Distribution}

To understand the bolometric luminosity of Mrk 817 at the time of our observation, we jointly fit the EPIC-pn spectrum from the long XMM-Newton observation and the six UVOT filters from the simultaneous Swift observation.  The best-fit model for the EPIC-pn spectrum was augmented with a multi-color disk blackbody component, \texttt{diskbb} (Mituda et al.\ 1984).  This component was not modified by the X-ray absorbers, but instead using the \texttt{redden} model within XSPEC.  The disk color temperature and flux normalization within \texttt{diskbb} were allowed to vary freely.  Only E(B-V) is set within \texttt{redden}; Following Kara et al.\ (2023), we set E(B-V) = 0.022.  

This model results in a broadly acceptable fit: $\chi^{2} = 953.7$ for $\nu = 856$ degrees of freedom.  The disk color temperature is measured to be $kT = 0.00200(2)$~keV, and the normalization is measured to be $K = 6.0^{+0.4}_{-0.1}\times 10^{11}$.  This normalization is defined as $K = (R_{in}/d_{10})^{2} cos\theta$, where $R_{in}$ is the inner radius in units of km and $d_{10}$ is the distance in units of 10~kpc.  For an inclination of $\theta = 19^{\circ}$ (Miller et al.\ 2020), this translates to a radius of $R = 1.07^{+0.07}_{-0.02}\times 10^{10}~km$, or $R = 185^{+13}_{-4}~GM/c^{2}$ for an assumed black hole mass of $M = 3.85\times 10^{7}~M_{\odot}$.  This would nominally indicate a truncated disk or a peak emission region that is significantly larger than the innermost stable circular orbit.  However, the disk continuum model should be regarded as fiducial; it is only implemented to characterize the flux in the optical and UV bands.  

Stripping all reddening and absorption away from the model, the implied bolometric luminosity is $L_{bol} = 8.7\pm 0.4\times 10^{43}~{\rm erg}~{\rm s}^{-1}$.  The implied ionizing luminosity, measured in the 13.6~eV to 13.6~keV band, is $L_{ion} = 7.2\pm 0.4 \times 10^{43}~{\rm erg}~{\rm s}^{-1}$.   The errors on these luminosity values were derived by varying the parameters of the continuum emission components within their respective 1$\sigma$ errors.  For a black hole mass $M = 3.85\times 10^{7}~M_{\odot}$  (Kara et al.\ 2021, Homayouni et al. 2023), this bolometric luminosity equates to an Eddington fraction of $\lambda = 0.016\pm0.001$; for the black hole mass of $M = 8.12^{+0.73}_{-0.67} \times 10^{7}~M_{\odot}$ measured by Lu et al.\ (2021), this equates to an Eddington fraction of $\lambda = 0.0083\pm 0.0007$.  Conservatively taking this range of masses as a systematic error that dominates statistical errors, the Eddington fraction of Mrk 817 during the XMM-Newton exposure was $\lambda = 0.008-0.016$.  

\subsection{A Complex X-ray Wind}

Figure 5 shows that the photoionized re-emission from the wind contributes significantly to the spectrum below 2~keV.   Particularly given the high column density values measured from these components, re-emission from the wind is expected.  Nominally, if the broadband SED can be modeled with perfect fidelity, the normalizations of the emission components (see Table 1) could be translated into distances and filling factors.   Even though we have performed basic SED modeling in order to determine the ionizing and bolometric luminosities at the time of our XMM-Newton observation, the photioionzation models that we have used still must assume a fixed spectral form as input.  In our fits, the emission normalizations in Table 1 are only meaningful in a relative sense.

As noted above, zones of paired absorption and re-emission were added to the model until the fit statistic did not improve significantly.  This procedure may not fully capture the degree to which each wind zone is required in the full spectral model.  We therefore removed zones from the best-fit model, and re-fit the data.  Removing zones 1--3 increases the fit statistic by $\Delta \chi^{2} = 143.3, 440.3, {\rm and} 327.7$, respectively.  Removing the unpaired emission component results in an increase of $\Delta \chi^{2} = 292.6$.  In each case, the change increases the number of degrees of freedom by $\Delta \nu = 4$.  Via a simple F test, each of these changes signal that every zone is required at far above the 8$\sigma$ level of confidence.  Consistent results are achieved if the zones are not removed, but if the columns and normalizations are instead frozen at minimum values.

The fact that column densities and ionization parameters are linked between absorption and emission components prevents an unbiased assessment of their significance by removing only one or the other.  However, in practice, the need for each component can be separately assessed by the confidence at which its key parameter excludes zero (column densities for the absorption zones, flux normalizations for their corresponding emission), and by understanding the spectra predicted by the ionization parameter.  For zones 1--3, dividing the best-fit column density by the $1\sigma$ minus-side error (e.g., ${\rm N}_{\rm H} / \sigma({\rm N}_{\rm H})$), gives values of 25.0, 14.4, and 7.9, respectively.  In this metric, zone 3 is the least constrained; moreover, its very high ionization, ${\rm log}\xi = 5.5^{+0.2}_{-0.1}$, means that even its high column fails to produce very strong absorption within the spectrum.  The characteristics of the fast outflows are therefore driven by zones 1 and 2.  

Key wind parameters, including radii, mass outflow rates, and kinetic power (in absolute terms, and relative to the bolometric luminosity and Eddington luminosity) are given in Table 2.  The ionization parameter formalism, $\xi = L_{ion}/nr^{2}$ (where $n$ is the number density of the gas, and $r$ is the radius between the central engine and gas cloud), can be rearranged to define a radius, $r^{2} = L_{ion}/n$.  Typically, the column density is measured but the number density is unknown, so writing $r = L_{ion}/N f \xi$ is pragmatic, where $f = \Delta r/r$ is a filling factor that can be orders of magnitude less than unity (for recent discussions, see Blustin et al.\ 2005 and Laha et al.\ 2016).  The ``absorption radius'' in Table 2 is an upper limit on the radius of the gas, obtained by assuming that $f = 1$.  The values obtained for zones 1 and 2, $R = 1.2^{+0.2}_{-0.08} \times 10^{4}~GM/c^{2}$ and $R = 1.0^{+0.2}_{-0.09}\times 10^{4}~GM/c^{2}$, are broadly consistent with the optical BLR or classical inner torus.  The absorption radius of zone 3 is much smaller, at $R = 35^{+3}_{-5}~GM/c^{2}$.  The ``wind radius'' of each zone is the radius at which the measured wind velocity is equal to the local escape velocity.  These radii are smaller for Zones 1 and 2, but may again function as upper limits, since the full wind velocity is likely higher than the velocity we measure in projection.  All three wind radii are consistent with values of $R \simeq 3 \;to\; 11\times 10^{2}~GM/c^{2}$.  The ratio of these radii can be regarded as constraint on the filling factor (see, e.g., Miller et al.\ 2016).  For zones 1 and 2, the filling factor is small, $f \sim 0.09$ (zone 1) and $f \sim 0.03$ (zone 2).  These values indicate that the wind may be clumpy or filamentary within these zones.  However, we note that comparable filling factors are found for the ``intermediate line region'' in NGC 4151, for which Crenshaw \& Kraemer (2007) derive $f = 0.04$.  For zone 3, we have assumed a filling factor of unity, since both radius estimates are very small.  

The mass outflow rates quoted in Table 2 were calculated via $\dot{M}_{wind} = \Omega \mu m_{p} n r^{2} v$, but substituting $L/\xi$ for $nr^{2}$ to give $\dot{M}_{wind} = \Omega \mu m_{p} (L/\xi) v$~ (where $\Omega$ is the covering factor and we assumed $\Omega = 2\pi$; $\mu$ is the mean atomic weight, taken to be $\mu = 1.23$, and $m_{p}$ is the mass of the proton).  This equation also implicitly assumes a volume filling factor of unity ($f=1$).  Table 2 also gives corrected mass outflow rates for zones 1 and 2 in parentheses, applying the filling factors derived by comparing plausible absorption and wind launching radii.  The kinetic power of each wind zone for both filling factors was then calculated via $L_{kin} = \frac{1}{2} \dot{M} v^{2}$.   Here again, the table gives both raw and corrected kinetic power values.

Even after correcting by data-driven filling factors, both zone 1 and zone 2 independently exceed the $L_{kin}/L_{bol} > 0.05$ threshold to shape host bulges identified by Di Matteo et al.\ (2005) and numerous subsequent treatments.  For zone 1, $L_{kin}/L_{bol} = 0.5^{+0.2}_{-0.07}$, and for zone 2, $L_{kin}/L_{bol} = 0.49^{+0.05}_{-0.04}$.   Thus, even discounting zone 3 ($L_{kin}/L_{bol} = 0.19^{+0.02}_{-0.03}$) because its column density and ionization are close to the model limits, the observed flow would still greatly exceed the theoretical threshold for fierce feedback.
For reference, Table 2 also quotes raw and corrected values for $L_{kin}/L_{Edd}$, assuming a black hole mass of $M = 8.12^{+0.73}_{-0.67}\times 10^{7}~M_{\odot}$ (Lu et al.\ 2021).  For the lower black hole mass of $M = 3.85\times 10^{7}~M_{\odot}$ (e.g., Kara et al.\ 2021, Homayouni et al.\ 2023), the values of $L_{kin}/L_{Edd}$ listed in Table 2 need to be multiplied by a factor of $\sim 2.1$.

The filling factors that we have derived have the merit of being data--driven, but they are imperfect.  It is possible that the true filling factor of each wind component is lower.  A very low limit on the filling factor might be determined by assuming that the winds are launched within the BLR, and utilizing the filling factor of cold obscuring clumps within that region.  The comparatively hot, ionized gas that we observe is more likely to pressure-confine such clumps, than to have a similarly small filling factor. Utilizing an eclipse of the central engine of NGC 6814 by a clump in the BLR, Gallo, Gonzalez, \& Miller (2021) report a characteristic size of $\Delta r = 1.3^{+0.17}_{-0.42}\times 10^{13}$~cm, at a distance of $r \sim 4.3\times 10^{15}$~cm, giving a filling factor of $C_{V} \simeq 0.003$.  Even if this filling factor is applied to the uncorrected values in Table 2, zone 2 still meets the 5\% threshold for fierce feedback.  

In order for $L_{kin}/L_{bol}$ to drop below the 5\% threshold, the filling factors must be lower than $f \simeq 0.009$ (zone 1), $f \simeq 0.003$ (zone 2), and $f \simeq 0.3$ (zone 3). For zones 1 and 2, the data-driven filling factors are well above these limits.  The limit for zone 2 equals the filling factor implied by eclipses in NGC 6814.  Future work may reveal even lower filling factors that do not satisfy the 5\% threshold, but all of the current estimates, using multiple techniques, point to a level of wind feedback in Mrk 817 that exceeds the threshold for significant impact on its host galaxy.

\section{Discussion}

We have made a detailed analysis of a moderately deep XMM-Newton/EPIC-pn spectrum of the Seyfert 1.2 AGN Mrk 817 in a very low flux state.  We find that the source luminosity is somewhat lower than that inferred in brighter states, but strong obscuration has driven most of the flux decrement.  Unlike the distant, neutral obscuration that shapes Seyfert-2 AGN, the data require a model with a combination of neutral and highly ionized absorption zones with coupled re-emission.  The ionized zones have projected blue-shifts of $v/c = 0.043^{+0.007}_{-0.003}$, $v/c = 0.079^{+0.003}_{-0.0008}$, and $v/c = 0.074^{+0.004}_{-0.005}$, signaling that they are components within a structured ultra-fast outflow.  For plausible, data-driven volume filling factors, this wind exceeds the theoretically predicted $L_{kin}/L_{bol} \geq 0.05$ threshold to reshape the host bulge (see Table 2).  Even if a lower filling factor is adopted based on recent eclipse studies (Gallo, Gonzalez, \& Miller 2021), the threshold is likely still met.  This inferred outflow is observed at an Eddington fraction of $\lambda \sim 0.008-0.016$ (depending on the assumed mass), potentially signaling that fierce feedback may not be limited to comparatively brief super-Eddington phases. 

AGN lifetimes are notoriously difficult to estimate, but typical numbers are in the range of $10^{7}$ -- $10^{9}$ years (Soltan 1982, Yu \& Tremaine 2002, Martini \& Weinberg 2001, Marconi et al. 2004).  However, it is likely that these estimates capture the total time span over which black holes eventually gain their mass, rather than the duty cycle within the growth span.  Noting that the light travel times across even large galaxies are only $10^{5}$ years, and comparing AGN with and without narrow lines excited by AGN activity, Schawinski et al.\ (2015) suggest that AGN phases may typically last $10^{5}$ years.  In this case, highly luminous AGN phases may be brief outbursts within total growth periods that are orders of magnitude longer.

Theoretical work by Novak, Ostriker, \& Ciotti (2011) also concludes that AGN phases are $10^{5}$~year bursts within $\sim10^{8}$ year spans.  In particular, those simulations predict rapid spikes to near-Eddington maxima, followed by relatively slow declines (qualitatively similar to outbursts in X-ray binaries).  If UFOs can be driven at substantially sub-Eddington luminosities, the time over which this feedback mode can operate may be longer by roughly an order of magnitude.   Many other well-known UFOs found in AGN, such as the strong X-ray outflow in PDS 456 (Nardini et al.\ 2015) and the X-ray and UV outflow in PG~1211$+$143 (e.g., Reeves et al.\ 2018, Kriss et al.\ 2018), are plausibly the result of quasi- or super-Eddington accretion within those systems.  The finding that UFOs and strong feedback can be driven in Mrk 817 at $\lambda \sim 0.008-0.016$ signals that such flows may not only have the power but also the {\em time} needed to reshape host galaxies.

It is interesting to explore how UFOs might be driven at substantially sub-Eddington mass accretion rates, when electron scattering pressure may be insufficient.  At low or moderate ionization levels, radiation pressure on specific UV transitions can drive fast winds.  This mechanism is likely responsible for the fast winds detected in broad absorption-line quasars (BALQSOs).  However, above ${\rm log}~\xi \geq 2$, few UV transitions remain, and this is no longer a viable wind driving mechanism (e.g., Proga et al.\ 2000).  The ionization levels measured in our best-fit model are much higher than this limit.  Recent studies of thermally driven winds, wherein outer regions of the accretion disk are heated to the local escape speed, show that large mass outflow rates are possible but that flow speeds are generally less than $v \leq 200~{\rm km}~{\rm s}^{-1}$ (Higginbottom et al. 2017).  Moreover, such winds can only be launched from large radii. 

Magnetic pressure from viscosity within the disk itself (e.g. Proga 2003) or poloidal field lines (e.g., Blandford \& Payne 1982) may play a role in launching winds in AGN.  In a comparison of slower X-ray winds and UFOs, Tombesi et al.\ (2013) find that magnetic driving is a viable explanation for the fastest, mostly highly ionized, and most powerful winds.  Within that work, a number of UFOs appear to exceed the 5\% threshold for strong feedback; in two cases, the inferred outflow power nominally exceeds the bolometric radiative luminosity.  These extreme cases may match the situation that we may have observed in Mrk 817.  Although slower winds may be driven by independent mechanisms, they may derive at least some of their power from magnetic driving.  The ``absorption measure distribution'' or AMD traces the distribution of wind properties with radius, and the shallow AMD measured in slow ``warm absorber'' winds in several Seyferts is also consistent with magnetic driving (e.g., Behar 2009).  Independently, careful studies of potential transverse velocities in NGC 4151 have concluded that magnetocentrifugal (Blandford-Payne) driving is required to explain some slow winds observed in NGC 4151 (Crenshaw \& Kraemer 2007).  The outer regions of the disks in AGN and stellar-mass black holes may differ, but their inner regions likely have key similarities, and high-resolution spectroscopy of stellar-mass black holes with Chandra suggests that magnetic pressure may also drive fast, ionized winds with kinetic powers that rival or exceed the radiative luminosity (e.g., King et al.\ 2012; Miller et al.\ 2015, 2016; Fukumura et al. 2017; Trueba et al.\ 2019; Del Santo et al. 2023).   In short, magnetic driving may be the most viable way to launch and power the winds that we have observed in Mrk 817.

In any AGN wind, the value of the filling factor (or, clumping factor) partly determines how much feedback the flow delivers into its host galaxy.  In the case of UFOs, this factor is directly related to whether or not the wind clears the threshold to reshape the bulge.  We have estimated the filling factor of the fast wind zones in Mrk 817 by comparing the absorption radii derived using a unity filling factor, with the radii at which the projected velocity of each zone would have the local escape speed.  The full velocity of each component is likely higher than its observed velocity, and in this case the true filling factors could be marginally smaller.  However, the total mass outflow rate and kinetic power of each component would more than compensate, since $\dot{M}_{wind} \propto v$ and $L_{kin} \propto v^{3}$.

Clumping is natural in low or moderately ionized winds driven by radiation pressure (see, e.g., Dannen et al.\ 2020, Waters et al.\ 2021), but those that we have detected in Mrk~817 are highly ionized.  Although the wind zones differ in ionization, their projected velocities are broadly similar, suggesting that the individual zones are more likely to be co-spatial.  It is less likely that the zones represent truly separate flows, separated by orders of magnitude in radius, and driven by distinct mechanisms.   The zones can only be co-spatial if the less ionized wind components have low filling factors, roughly given by the radio of the ``wind'' and ``absorption'' radii (see Table 2).

Whereas it is common for inferred mass outflow rates to exceed inferred accretion rates, it may seem less natural for the kinetic power in a wind to exceed the radiative luminosity.  However, there is a broad parameter space wherein this is easily feasible, especially in fast, highly ionized UFOs that may be magnetically driven.  Starting with the usual expression for the mass outflow rate, $\dot{M} = \Omega~m_{p}vnr^{2}$, substituting the ionization parameter formalism to arrive at $\dot{M} = \Omega m_{p}v(L_{ion}/\xi)$, one can then write the kinetic power as $L_{kin} = \frac{1}{2}\Omega~m_{p}v^{3}(L_{ion}/\xi)$.  If $L_{ion} \simeq L_{bol}$, then $L_{kin}/L_{bol} \simeq \frac{1}{2}\Omega m_{p}v^{3}/\xi$.  For $\xi \geq 2/\Omega$ and $v \geq 10^{8}~{\rm cm}~{\rm s}^{-1}$ (just $10^{3}~{\rm km}~{\rm s}^{-1}$), it is then natural that $L_{kin}/L_{bol} \geq 1$.  In Seyferts, $\Omega\simeq 0.5$ is often assumed based on the distribution of unobscured and obscured sources, so the condition on $\xi$ is easily met in ionized outflows.  Even if it is necessary to account for an additional clumping factor $f$ within the $\Delta r$ band where the wind is launched, changing the condition on ionization to be $\xi \geq 2/\Omega f$, the high ionization of UFOs could overcome small values of $f$.  The value of $f$ cannot be too small, of course, or the product $\Omega f$ would make the detection of outflows very rare.

The observations that we have analyzed in this work were made 1.5 years after the observations that formed the basis of the recent AGN-STORM2 campaign (see, e.g., Kara et al.\ 2021, Homayouni et al.\ 2023).  That campaign confirmed and extended evidence of a variable, ionized, partial covering wind reported by Miller et al.\ (2021).  No UFOs were reported in X-rays, but this could be an artifact of lower spectral sensitivity in the related XMM-Newton spectra, or the limited spectral resolution of the related NICER monitoring program (Partington et al.\ 2023).  It is also possible that our observations simply captured a less dynamic phase with clearer outflow structure, at least in X-rays.  The AGN-STORM2 campaign did measure UV line shifts of $v = -3700~{\rm km}~{\rm s}^{-1}$ to $v = -5500~{\rm km}~{\rm s}^{-1}$, corresponding to $v/c = -0.012$ and $v/c = -0.018$.  These outflows are only a factor of a few slower than those detected in our deep XMM-Newton exposure (see Table 1), and it is possible that a slightly higher column density in a more ionized component would have revealed the X-ray component of the wind in the AGN-STORM2 campaign.  In this case, strong outflows representing fierce feedback would be a common, high-duty-cycle feature of Mrk 817.

The best-fit model detailed in Table 1 is complex, and the fit is not formally acceptable.  The model framework that we have applied likely captures the key facets of the complex outflow and distant reflection in Mrk 817, but it is necessarily incomplete.  The characteristics of the fast wind absorption are driven by zones 1 and 2, while zone 3 is not well constrained by the data.  Especially given its very high ionization, the contributions from zone 3 might be at least partially described in terms of scattered light.  Separately, our models assume that absorbing gas at some radius $R$ on the near side of the central engine does not cover any of the emission from gas at $R$ on the far side of the black hole.  The framework also assumes that no absorption zone covers emission from any other zone.  These assumptions may be largely justified even for polar outflows given the low inclination at which Mrk 817 may be viewed.  However, interactions between zones are plausible and would serve to cause small changes in the total spectral model.  Future work on these observations can improve upon such shortcomings, and may arrive at a more complete picture of the outflow.  

The recent launch of XRISM (Tashiro et al.\ 2020) marks the best opportunity to test our model, and to explore the nature of variable accretion flows and strong winds in Seyfert AGN (see, e.g., Gallo, Miller, \& Costantini 2023).   The 5~eV resolution expected from the Resolve calorimeter spectrometer will greatly facilitate the detection and characterization of fast X-ray outflows.  To illustrate this, we simulated a 100~ks Resolve spectrum of Mrk 817 using publicly available responses, assuming the best-fit model detailed in Table 1.  The spectra are shown in Figure 7, in a manner that corresponds to Figure 4.  XRISM is smaller than XMM-Newton, particularly at low energy but also in the Fe~K band;  however, the sharp resolution of the spectra concentrates flux in the absorption lines associated with the fastest outflows.  In Mrk 817 and similarly variable sources, then, monitoring studies with XRISM can trace the evolution and duty cycle of X-ray winds.

We thank the anonymous referee for comments that improved this manuscript.  We also thank the PI of XMM-Newton, Norbert Schartel, and the PI of NuSTAR, Fiona Harrison, for granting these observations, and their mission teams for executing them.


\begin{deluxetable}{c|cc}
\tablecaption{Model Parameters}
\label{table:results}
\tablewidth{\textwidth} 
\tabletypesize{\scriptsize}
\tablehead{
\colhead{Model Component} & \colhead{Parameter} & \colhead{Value}
}
\startdata\\
$\texttt{TBabs}$ & $N_{\rm H} \: [\rm \times 10^{22} \: cm^{-2}]$ & $1.06 \times 10^{-2}*$ \\\hline
$\texttt{tbpcf}$ & $N_{\rm H} \: [\rm \times 10^{22} \: cm^{-2}]$ & $29^{+3}_{-2}$ \\
 & $\rm CvrFract$ & $0.80^{+0.03}_{-0.05}$ \\ 
 & $\rm Redshift$ & $3.14 \times 10^{-2}*$ \\\hline
$\texttt{abs<1>}$ & $N_{\rm H} \: [\rm \times 10^{22} \: cm^{-2}]$ & $25^{+4}_{-1}$ \\
 & $\rm \log(\xi)\;[\rm erg\;cm\;s^{-1}]$ & $3.31^{+0.04}_{-0.01}$ \\
 & $\rm Redshift$ & $-1.1^{+0.7}_{-0.3} \times 10^{-2}$ \\\hline
 $\texttt{abs<2>}$ & $N_{\rm H} \: [\rm \times 10^{22} \: cm^{-2}]$ & $13^{+2}_{-0.9}$ \\
 & $\rm \log(\xi)\;[\rm erg\;cm\;s^{-1}]$ & $3.66^{+0.02}_{-0.02}$ \\
 & $\rm Redshift$ & $-4.8^{+0.3}_{-0.08} \times 10^{-2}$ \\\hline
$\texttt{abs<3>}$ & $N_{\rm H} \: [\rm \times 10^{22} \: cm^{-2}]$ & $55^{+3}_{-7}$ \\
 & $\rm \log(\xi)\;[\rm erg\;cm\;s^{-1}]$ & $5.5^{+0.2}_{-0.1}$ \\
 & $\rm Redshift$ & $-4.2^{+0.4}_{-0.5} \times 10^{-2}$ \\\hline
$\texttt{cutoffpl}$ & $\rm PhoIndex$ & $1.8^{+0.09}_{-0.1}$ \\
 & $\rm HighECut\;[keV]$ & $100.0*$ \\
 & $\rm norm$ & $3.1^{+0.7}_{-0.6} \times 10^{-3}$ \\\hline 
$\texttt{emi<1>}$ & $N_{\rm H} \: [\rm \times 10^{22} \: cm^{-2}]$ & $25^{+3}_{-2}\dagger$ \\
 & $\rm \log(\xi)\;[\rm erg\;cm\;s^{-1}]$ & $3.31^{+0.04}_{-0.01}\dagger$ \\
 & $\rm Redshift$ & $3.14 \times 10^{-2}*$ \\
 & $\rm norm$ & $<\;2.72\times10^{-6}$ \\\hline
 $\texttt{emi<2>}$ & $N_{\rm H} \: [\rm \times 10^{22} \: cm^{-2}]$ & $13^{+2}_{-0.9}\dagger$ \\
 & $\rm \log(\xi)\;[\rm erg\;cm\;s^{-1}]$ & $3.66^{+0.02}_{-0.02}\dagger$ \\
 & $\rm Redshift$ & $3.14 \times 10^{-2}*$ \\
 & $\rm norm$ & $1.8^{+0.2}_{-0.2} \times 10^{-4}$ \\\hline
 $\texttt{emi<3>}$ & $N_{\rm H} \: [\rm \times 10^{22} \: cm^{-2}]$ & $55^{+3}_{-7}\dagger$ \\
 & $\rm \log(\xi)\;[\rm erg\;cm\;s^{-1}]$ & $5.5^{+0.2}_{-0.1}\dagger$ \\
 & $\rm Redshift$ & $3.14 \times 10^{-2}*$ \\
 & $\rm norm$ & $8^{+2}_{-2}\times 10^{-4}$ \\\hline
 $\texttt{emi<4>}$ & $N_{\rm H} \: [\rm \times 10^{22} \: cm^{-2}]$ & $5.0*$ \\
 & $\rm \log(\xi)\;[\rm erg\;cm\;s^{-1}]$ & $-0.98^{+0.08}_{-0.08}$ \\
 & $\rm Redshift$ & $3.14 \times 10^{-2}*$ \\
 & $\rm norm$ & $2.2^{+0.6}_{-0.3} \times 10^{-2}$ \\\hline
$\texttt{pexmon}$ & $\rm PhoIndex$ & $1.8^{+0.09}_{-0.1}\dagger$ \\
 & $\rm foldE\;[keV]$ & $100.0*$ \\
 & $\rm rel\_refl$ & $-0.20^{+0.03}_{-0.04}$ \\
 & $\rm Redshift$ & $3.14 \times 10^{-2}*$ \\
 & $\rm abund$ & $1.0*$ \\
 & $\rm Fe\_abund$ & $1.0*$ \\
 & $\rm Incl\;[^{\circ}]$ & $19.0*$ \\
 & $\rm norm$ & $3.1^{+0.7}_{-0.6} \times 10^{-3}\dagger$ \\\hline
 & $\chi^2 / \nu$ & $ 2089.39/1746 \;=\;1.20 $ \\\hline 
\enddata

\tablecomments{Measured values and $1\sigma$ errors for the parameters of our best-fit model.  The very complex spectrum of Mrk 817 requires neutral partial-covering absorption, three zones of paired photoionized absorption plus re-emission.  These zones were modeled using XSTAR tables (please see the text for details.)  The absorbers all affect a direct continuum consisting of a cut-off power-law.  Distant, neutral reflection is also required and modeled using ``pexmon.''  The photoionized emission and distant reflection components are not covered by the absorbers.  An additional low-ionization photoionized emission component is required that may correspond to diffuse gas in the, e.g., NLR.  Values marked with $*$ are frozen.  Emission parameters marked with $\dagger$ are tied to their associated absorption parameters.  All velocity shifts are reported relative the host, which is observed at a redshift of $z=0.03145$ (Strauss \& Huchra 1988). }
\end{deluxetable}

\begin{deluxetable}{c|ccc}
\tablecaption{Wind Parameters}
\label{table:wind}
\tablewidth{\textwidth} 
\tabletypesize{\scriptsize}
\tablehead{
\colhead{ } & \colhead{\texttt{abs<1>}} & \colhead{\texttt{abs<2>}} & \colhead{\texttt{abs<3>}}
}
\startdata\\
$\texttt{Velocity $10^{4} \rm (km\;s^{-1})$}$ & $-1.3^{+0.2}_{-0.1}$ & $-2.37^{+0.09}_{-0.02}$ & $-2.2^{+0.1}_{-0.2}$ \\\hline
$\texttt{Velocity (c)}$ &$-0.043^{+0.007}_{-0.003}$ & $-0.079^{+0.003}_{-0.0008}$ & $-0.074^{+0.004}_{-0.005}$ \\\hline
$\texttt{Absorption Radius $\rm (GM\;c^{-2})$}$ & $1.2^{+0.2}_{-0.08} \times 10^{4}$ & $1.0^{+0.2}_{-0.09} \times 10^{4}$ & $35^{+3}_{-5}$ \\\hline
$\texttt{Wind Radius $\rm (GM\;c^{-2})$}$ & $1.1^{+0.3}_{-0.1} \times 10^{3}$ & $320^{+30}_{-30}$ & $370^{+40}_{-50}$ \\\hline
$\texttt{Mass Outflow Rate (g~${\rm s}^{-1}$)}$ & $6^{+1}_{-0.5} \times 10^{26}$ & $4.8^{+0.3}_{-0.3} \times 10^{26}$ & $6.5^{+0.6}_{-0.6} \times 10^{24}$ \\\newline & $(5.4^{+0.9}_{-0.5} \times 10^{25})$ & $(1.5^{+0.1}_{-0.09} \times 10^{25})$ & $ $ \\\hline
$\texttt{Mass Outflow Rate ($\rm M_{\odot}/yr$)}$ & $9^{+2}_{-0.8}$ & $7.7^{+0.5}_{-0.4}$ & $0.104^{+0.009}_{-0.009}$ \\\newline & $(0.9^{+0.1}_{-0.08})$ & $(0.24^{+0.02}_{-0.01})$ & $ $ \\\hline
$\texttt{Kinetic Power \rm (${\rm erg}\; \rm s^{-1}$)}$ & $5^{+1}_{-0.7} \times 10^{44}$ & $1.4^{+0.1}_{-0.08} \times 10^{45}$ & $1.6^{+0.2}_{-0.2} \times 10^{43}$ \\\newline & $(5^{+1}_{-0.6} \times 10^{43})$ & $(4.3^{+0.4}_{-0.3} \times 10^{43})$ & $ $ \\\hline
$\texttt{$L_{kin}$ / $L_{bol}$}$ & $6^{+2}_{-1}$ & $16^{+2}_{-2}$ & $0.19^{+0.02}_{-0.03}$ \\\newline & $(0.5^{+0.2}_{-0.07})$ & $(0.49^{+0.05}_{-0.04})$ & $ $ \\\hline
$\texttt{$L_{kin}$ / $L_{Edd}$}$ & $0.05^{+0.01}_{-0.008}$ & $0.13^{+0.02}_{-0.01}$ & $1.6^{+0.2}_{-0.2} \times 10^{-3}$ \\\newline & $(4^{+1}_{-0.7} \times 10^{-3})$ & $(4.2^{+0.5}_{-0.4} \times 10^{-3})$ & $ $ \\\hline
\enddata

\tablecomments{Estimated values and uncertainties for properties of each of the three wind zones.  For each zone, the ``absorption radius'' is the photoionization radius derived via $r = L/N\xi$, implicitly assuming a volume filling factor of unity.  The ``wind radius'' is the radius at which the measured velocity shift equals the local escape speed.
The absorption and wind radii are given in units of gravitational radii ($\rm GMc^{-2}$) assuming a black hole mass of $M = 8.12^{+0.73}_{-0.67} \times 10^{7}~M_{\odot}$ (Lu et al.\ 2021).  The kinetic power to Eddington luminosity ratio assumes the same mass.  A bolometric luminosity ($L_{bol}$) of $\rm 8.7 \times 10^{43}\; erg\; s^{-1}$ was assumed given the SED.  The mass outflow rate and kinetic power in each zone are quoted assuming both a volume filling factor of unity and, in parentheses, assuming the calculated filling factor for zones 1 and 2; please see the text for a related discussion.}
\end{deluxetable}

\clearpage

\begin{figure*}
    \centering
    \includegraphics[width= 0.8\textwidth]{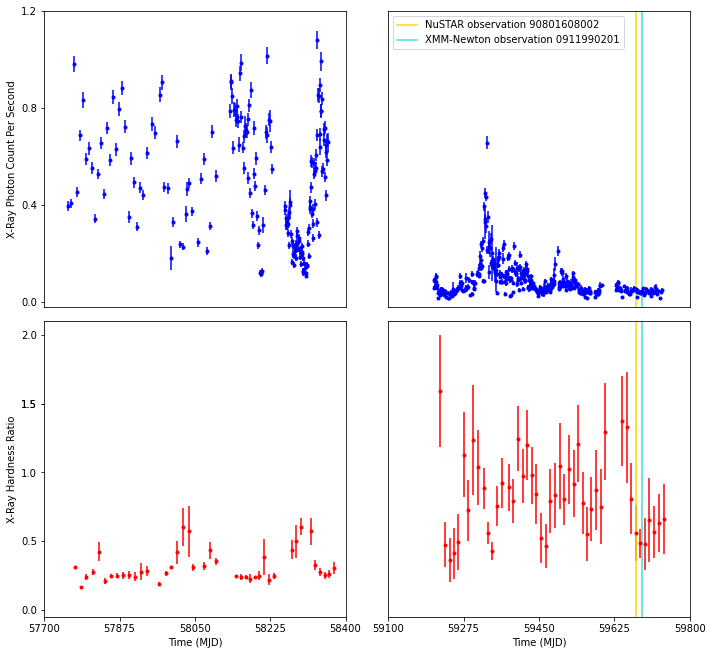}
    \caption{The recent Swift/XRT light curve is shown in the top two panels, and the X-ray hardness ratio is shown in the two panels below. Data are plotted for two 700-day time intervals separated by a 700-day gap in monitoring. The top left-hand panel shows extreme variability, likely owing to a combination of intrinsic variability and strong intermittent disk winds. The top right-hand panel shows the deep low-flux state that we have explored with NuSTAR and XMM-Newton. The bottom right-hand panel shows the observed increase in X-ray hardness, as expected from the absorption in the soft X-ray band by disk winds. Each data point in the hardness ratio represents the average hard photon count (2.0-10.0 keV) divided by the average soft photon count (0.3-2.0 keV) in a ten day period.  The vertical yellow and cyan lines indicate the time of the NuSTAR and XMM-Newton exposures that are analyzed in this paper.}
    \label{fig:swlc}
\end{figure*}

\begin{figure*}
    \centering
    \includegraphics[width= 0.5\textwidth]{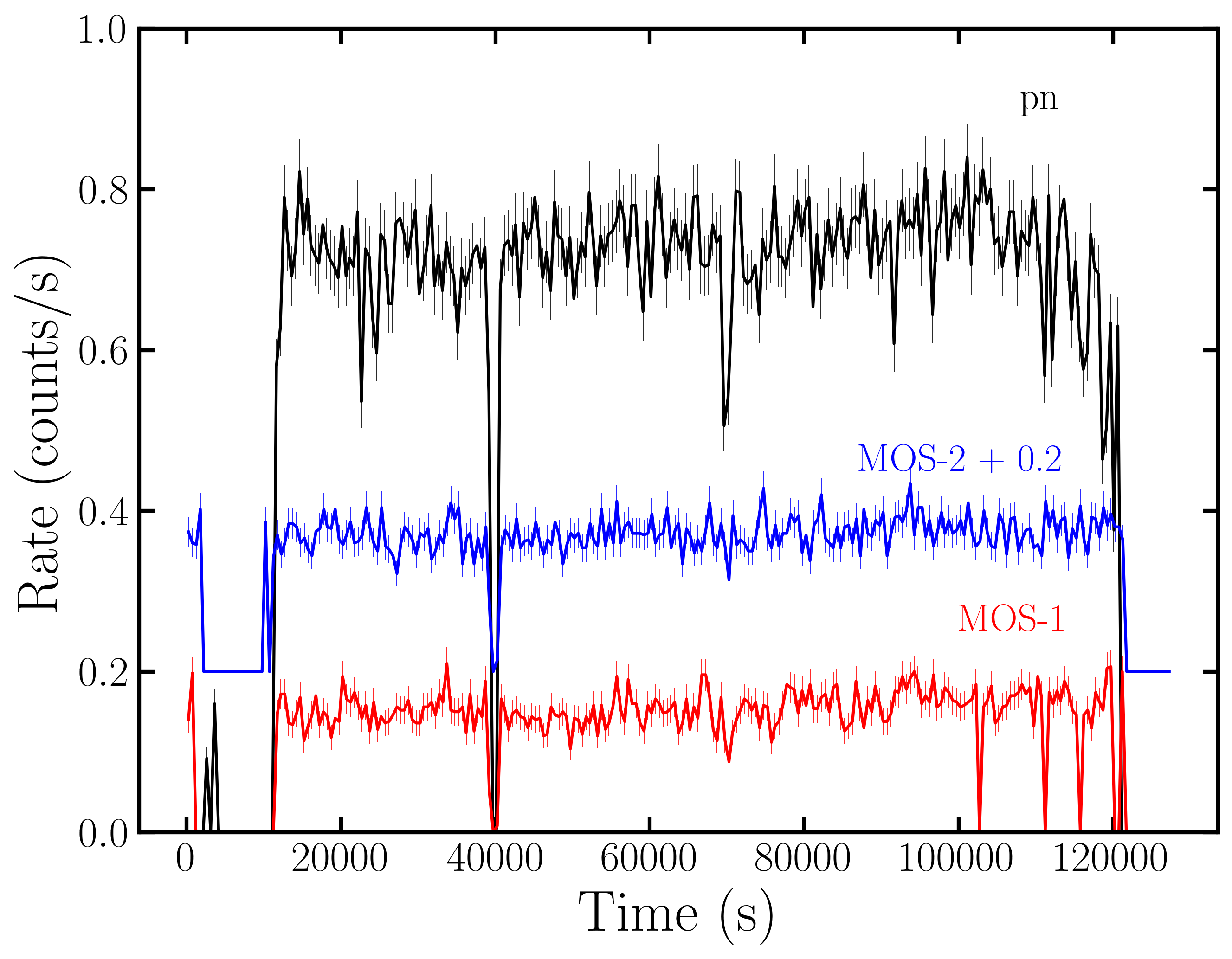}
    \caption{The source light curves from the three EPIC cameras aboard XMM-Newton, over the full spectral fitting band (0.3--10.0~keV).  The pn light curve is shown in black, the MOS-1 light curve in red, and the MOS-2 light curve in blue. 
 Note the MOS-2 light curve is artificially shifted by $+$0.2 counts per second for visual clarity.  The data were binned to have a resolution of 1000~s.  Periods of zero flux are due to the exclusion of instrumental flaring.  The zero point is measured relative to the start time of the MOS-1 exposure, which slightly preceded the start of the pn exposure.}
    \label{fig:xmmlc}
\end{figure*}

\clearpage

\begin{figure*}
    \centering
    \includegraphics[width= 0.33\textwidth,angle=-90]{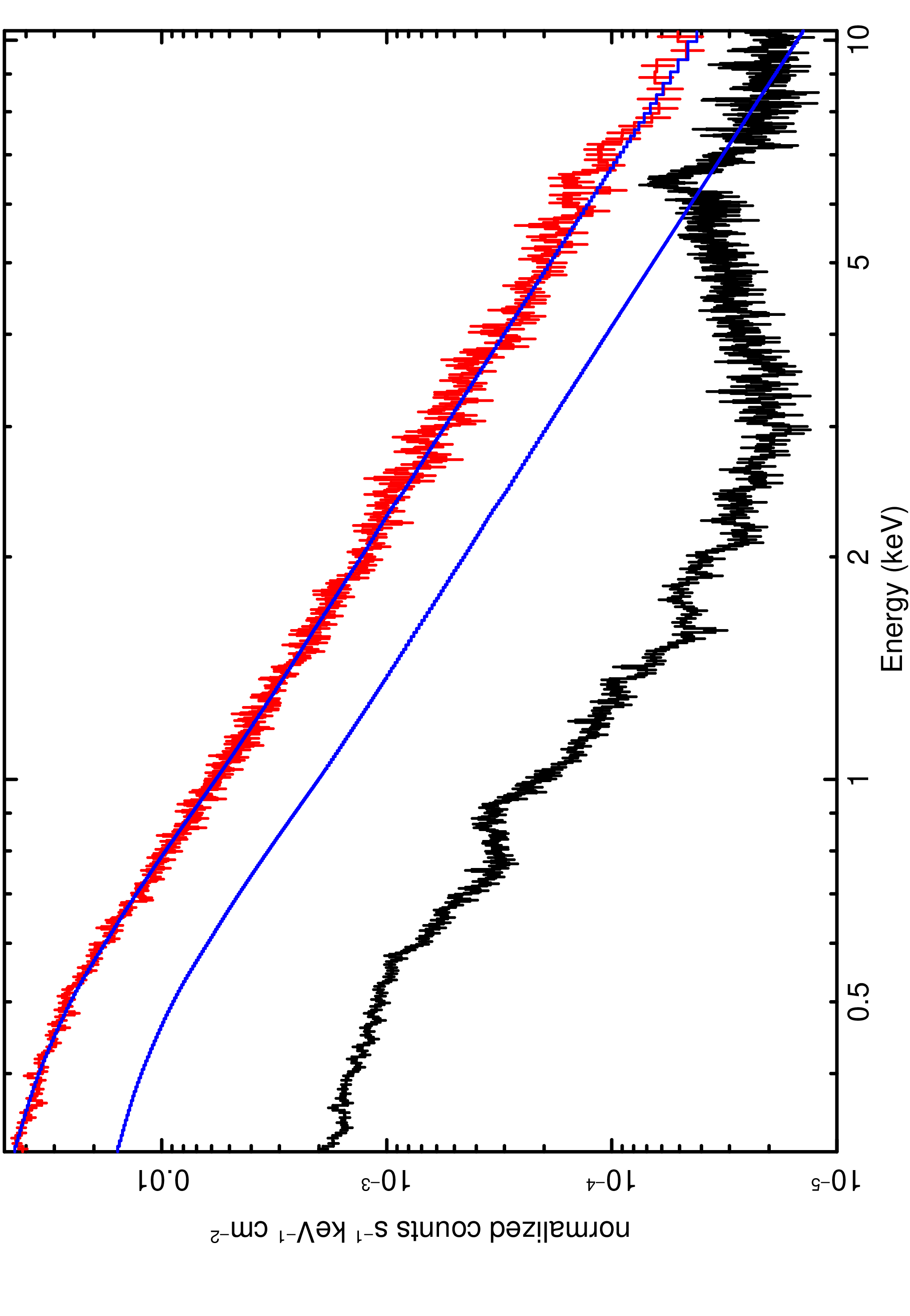}
    \includegraphics[width= 0.33\textwidth,angle=-90]{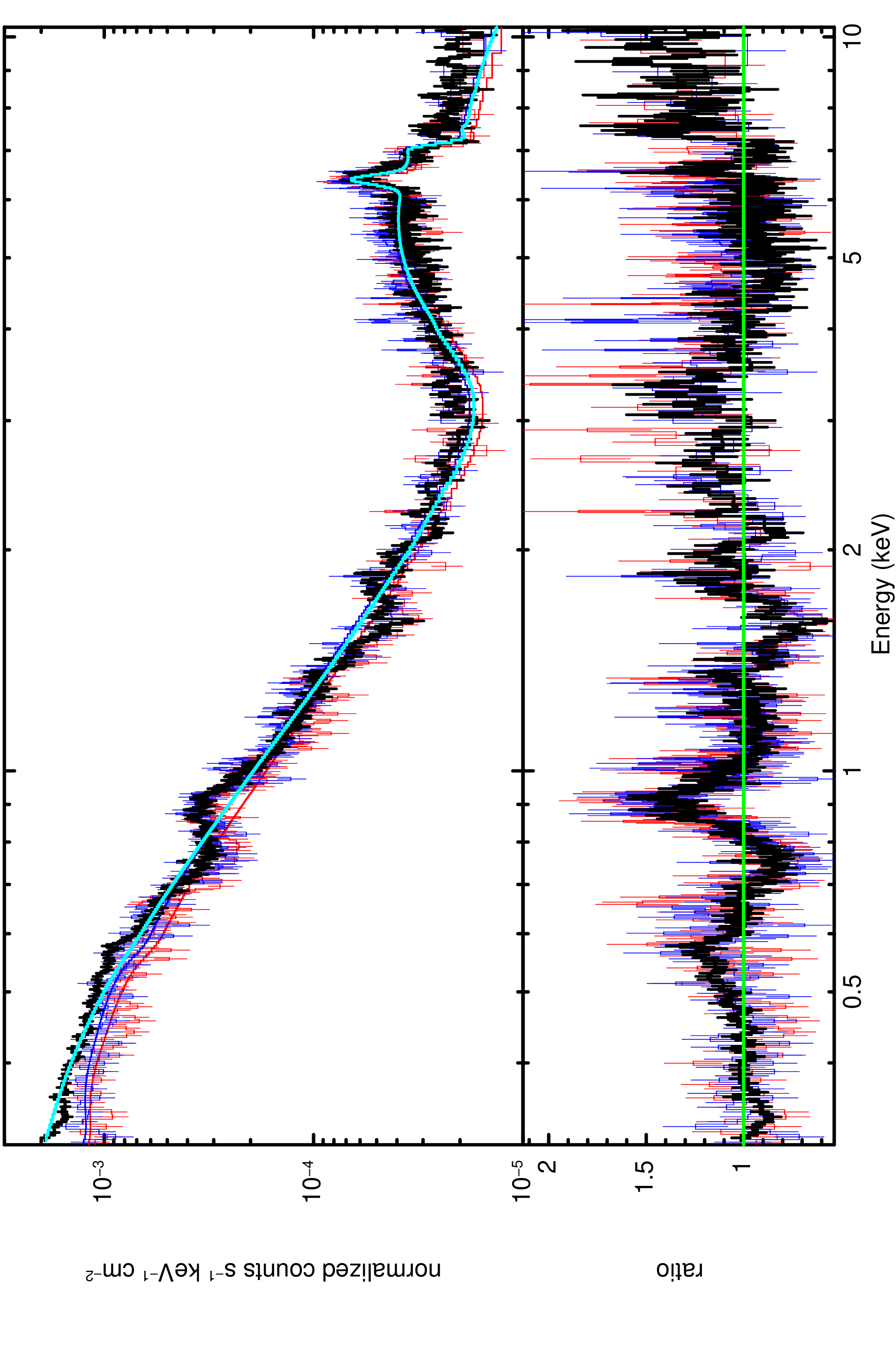}
    \caption{LEFT: XMM-Newton/EPIC-pn spectra of Mrk 817 in 2009 (red) and 2022 (black).  The data are shown in the host frame and are binned for visual clarity.  A fiducial, unabsorbed, broken power-law model (shown in blue) was fit to the spectrum in 2009, and then shifted to match the 2022 spectrum in the 7--10~keV band.  The 2022 spectrum is clearly marked by extremely strong, broad absorption and (re-)emission.  RIGHT: The 2022 EPIC-pn (black), MOS-1 (red), and MOS-2 (blue) spectra of Mrk 817, jointly fit with a model consisting of neutral partial covering of a cut-off power-law direct continuum, and neutral distant reflection (in XSPEC terminology, $tbpcf\times cutoffpl + pexmon$).  This model clearly fails to account for strong absorption and emission lines across the full pass band, pointing to the need for more complex, ionized absorption.}
    \label{fig:comp}
\end{figure*}

\begin{figure*}
    \centering
    \hspace{-0.4in}
    \includegraphics[width= 0.33\textwidth,angle=-90]{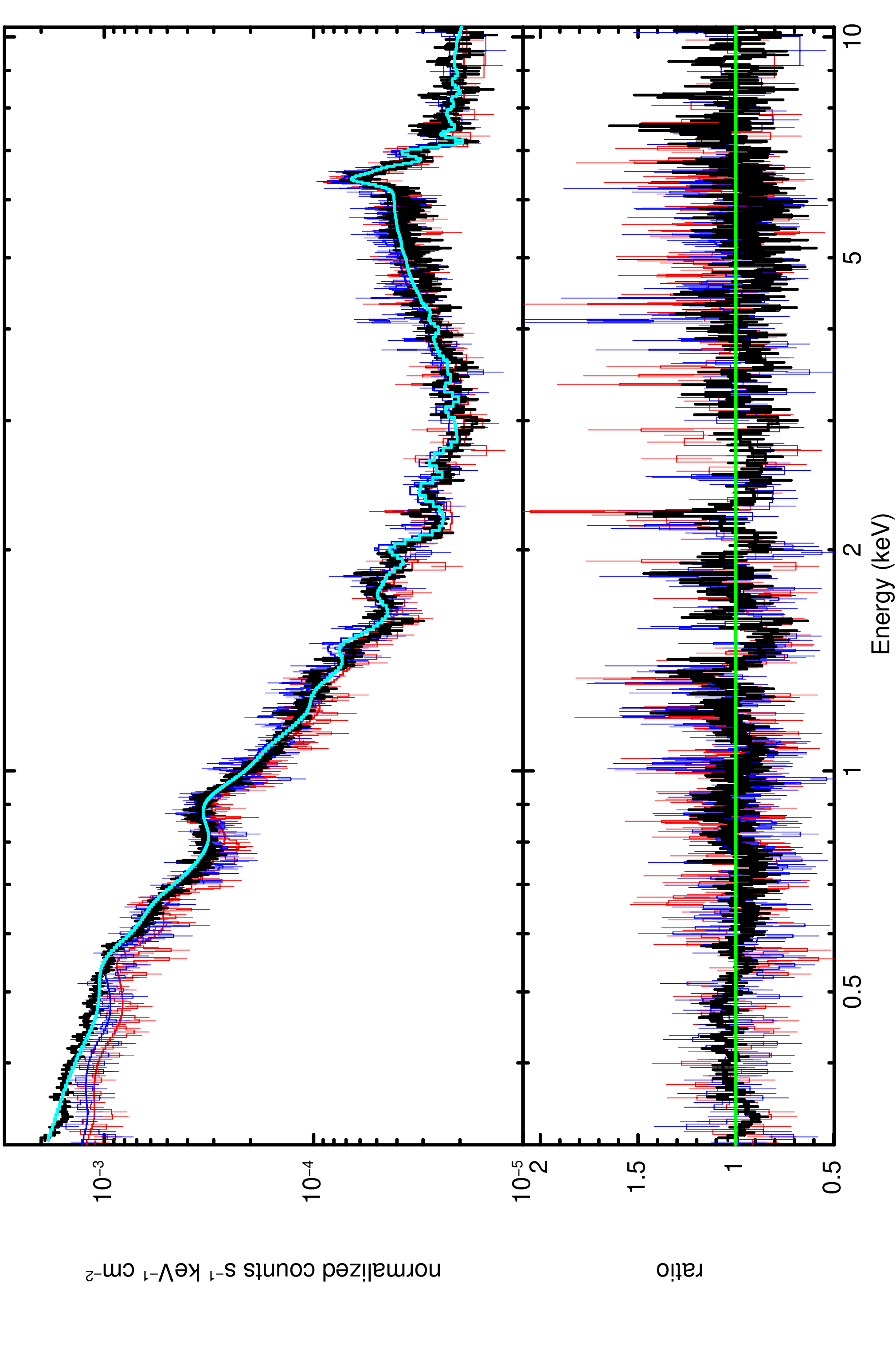}
        \includegraphics[width= 0.33\textwidth,angle=-90]{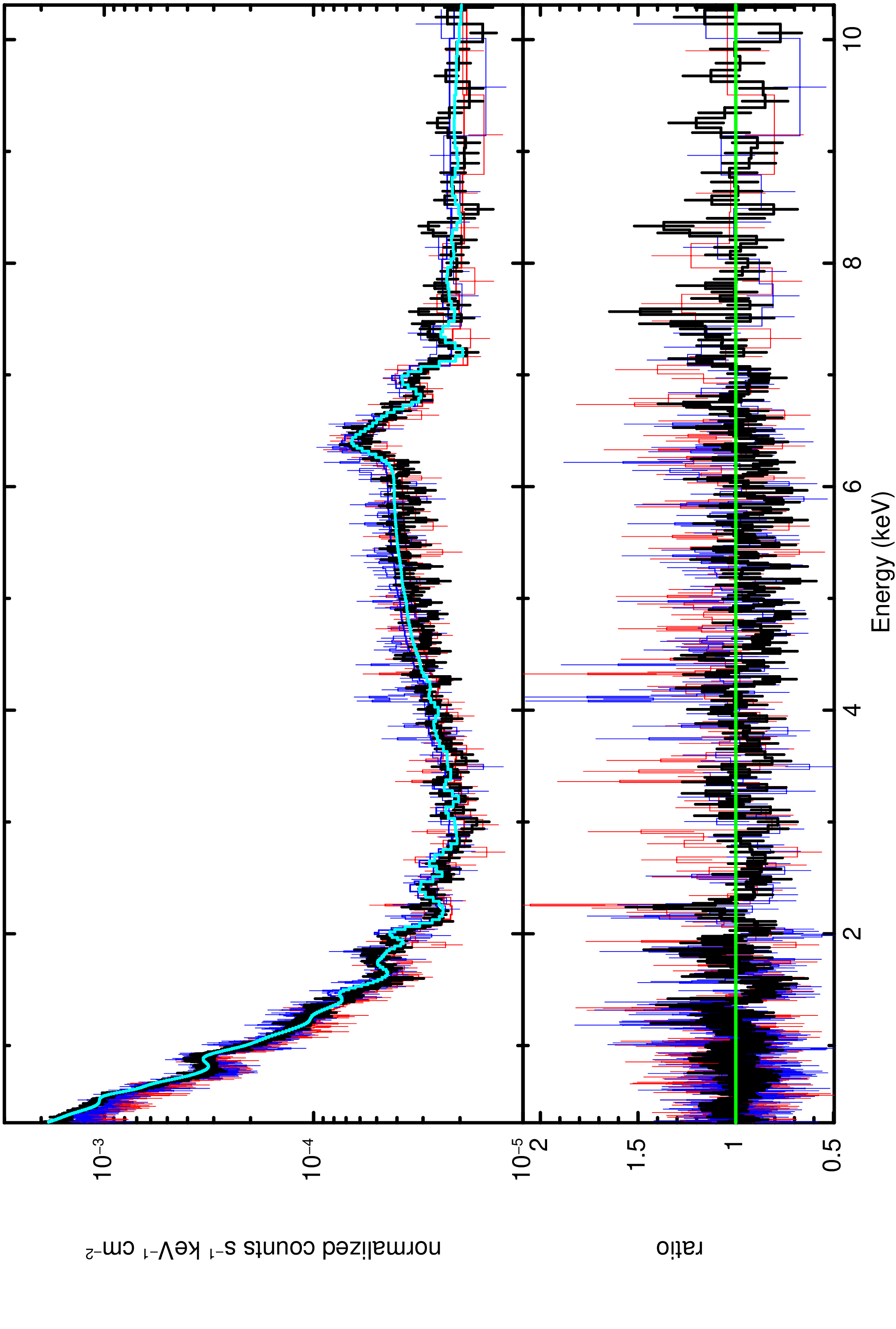}
    \caption{The best-fit model for the EPIC spectrum of Mrk 817 in its highly obscured state, and the associated data/model ratio.  The pn spectrum is shown in black, the MOS-1 spectrum in red, and the MOS-2 spectrum in blue.
    The left and right panels show log and linear energy scales, respectively.  The model includes neutral partial covering absorption, three zones of paired photoionized absorption and re-emission, and distant, neutral reflection.  The absorption affects a direct continuum consisting of a cut-off power-law.  The three ionized absorbers are part of an ultra-fast outflow, with projected velocities ranging between $v/c =$0.05--0.09.  This is likely best exemplified by the Fe XXVI line shifted from 6.97~keV to 7.25~keV.  The data were binned for visual clarity.  See Table 1 and the text for additional details.  }
    \label{fig:best}
\end{figure*}

\begin{figure*}
    \centering
    \hspace{-0.4in}
    \includegraphics[width= 0.4\textwidth,angle=-90]{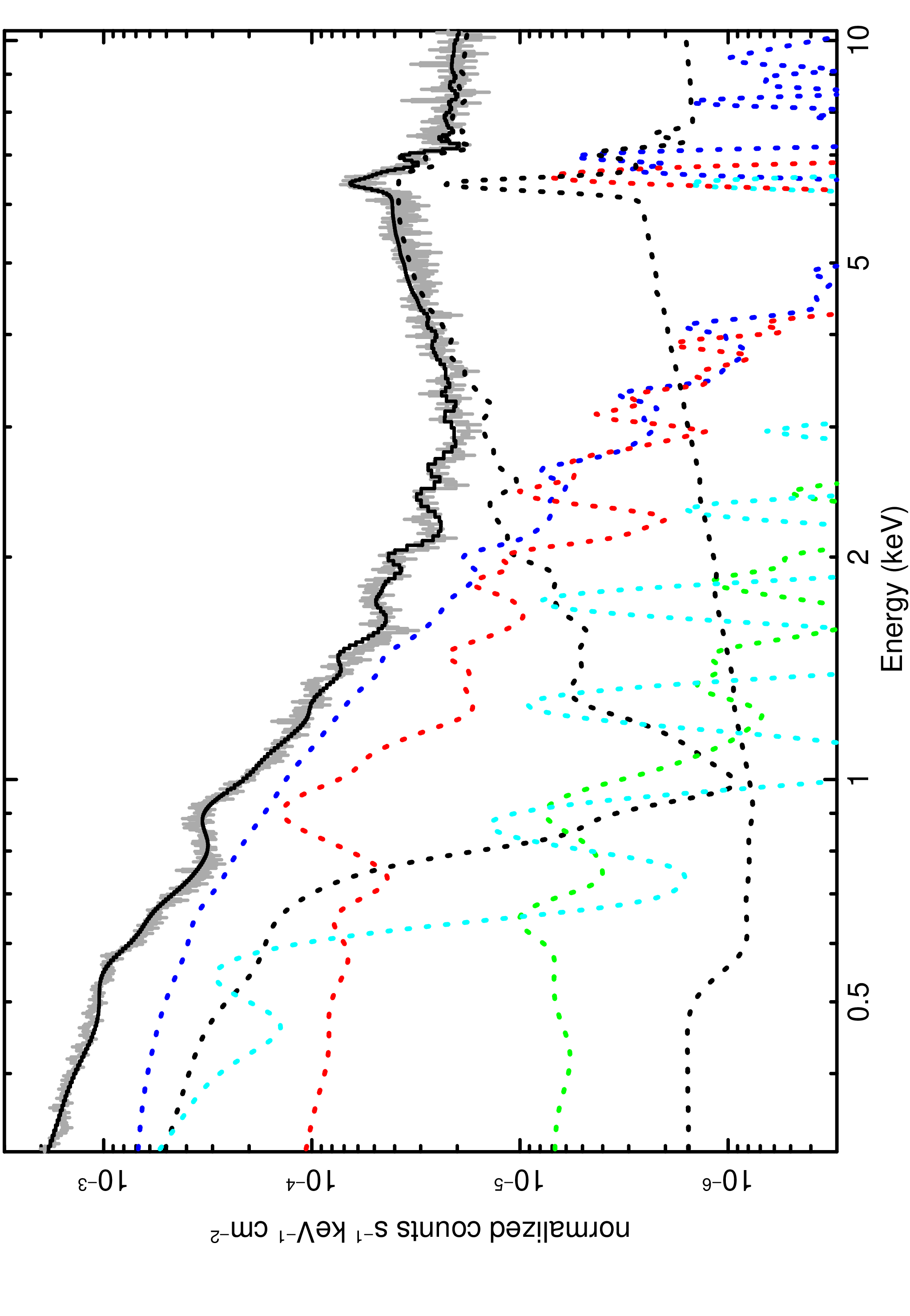}
    \caption{The best-fit model for the XMM-Newton spectra is shown here, including additive components.  For simplicity, only the pn spectrum is shown in gray, and the total best-fit model is shown in solid black.  Dashed black: the heavily absorved direct continuum (higher flux), and distant reflected flux.  Dashed red: the unpaired additive wind re-emission component, potentially from the NLR.  Dashed cyan, green, and blue (respectively): wind re-emission from zones 1, 2, and 3 (respectively).}
    \label{fig:lin}
\end{figure*}



\begin{figure*}
    \centering
    \hspace{-0.4in}
    \includegraphics[width= 0.4\textwidth,angle=-90]{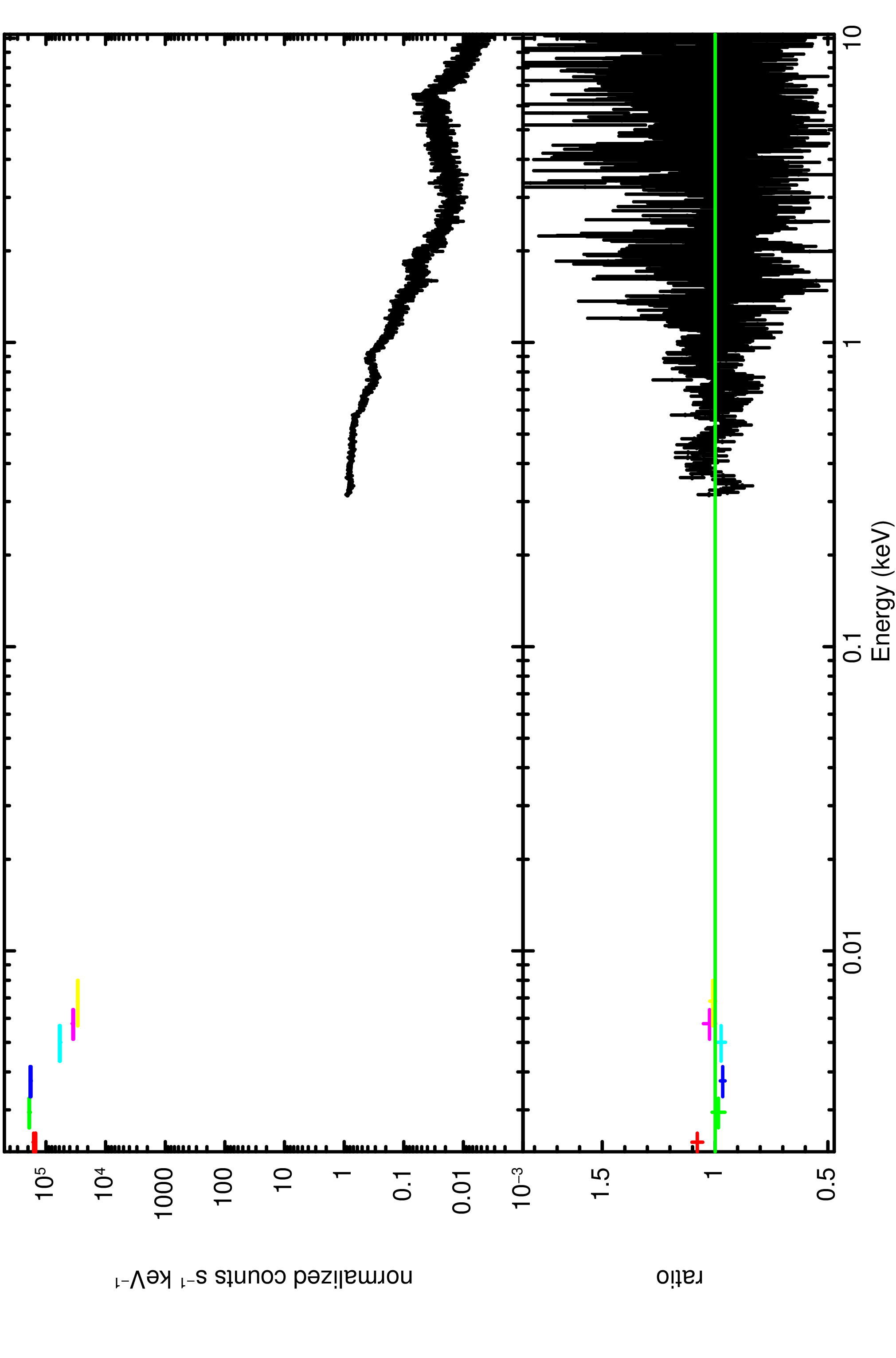}
    \caption{The broad spectral energy distribution (SED) of Mrk 817.  The spectrum shown in black is the EPIC-pn spectrum shown in Figures 2--5, fit with the model shown in Figures 4 and 5 and detailed in Table 1.  The six optical and UV flux points are Swift/UVOT filter fluxes derived from a simultaneous observation.  The in order of ascending energy, the B, V, U, UVW1, UVM2, and UVW2 filters are shown in red, green, blue, cyan, magenta, and yellow, respectively.  The fit to the EPIC-pn spectrum was extended to longer wavelengths with the addition of a reddened disk blackbody function.  When the broad unabsorbed flux is translated into a bolometric luminosity, an Eddington fraction of $\lambda = 0.008-0.016$ is implied (the uncertainty reflects a range of reported black hole mass values).}
    \label{fig:sed}
\end{figure*}

\begin{figure*}
    \centering
    \hspace{-0.4in}
   \includegraphics[width= 0.55\textwidth,angle=0]{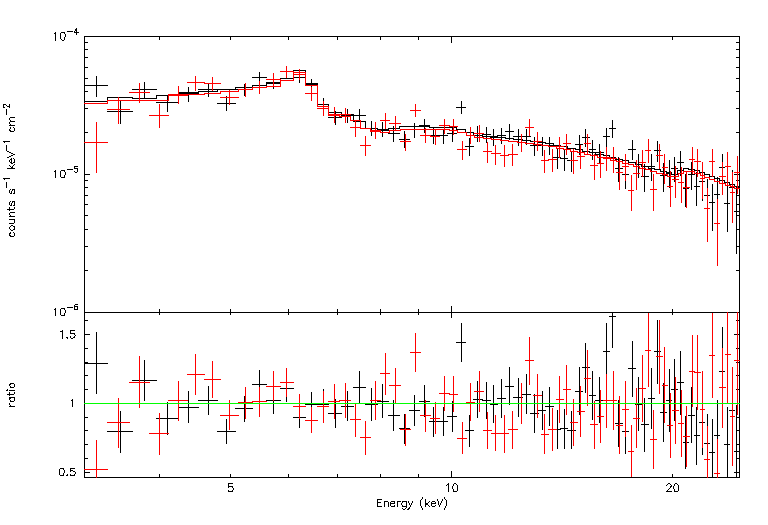}
    \caption{The joint best-fit models for the NuSTAR FPMA (black) and FPMB (red) spectra of Mrk 817 in the range 3-25 keV. The model fits well despite its relative complexity compared to the spectral resolution of the NuSTAR data.}
    \label{fig:nustar1}
\end{figure*}

\begin{figure*}
    \centering
    \hspace{-0.4in}
    \includegraphics[width= 0.33\textwidth,angle=-90]{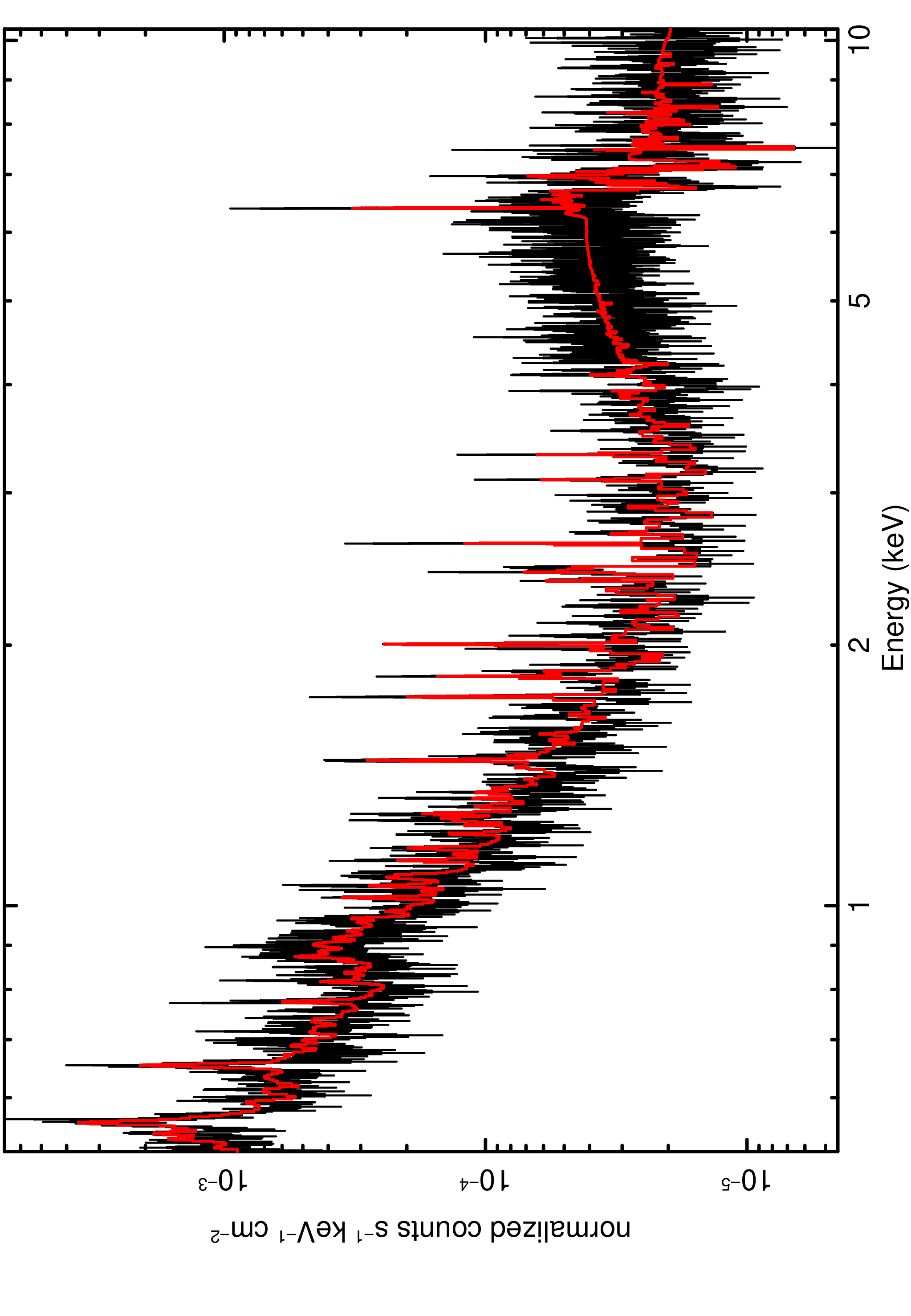}
        \includegraphics[width= 0.33\textwidth,angle=-90]{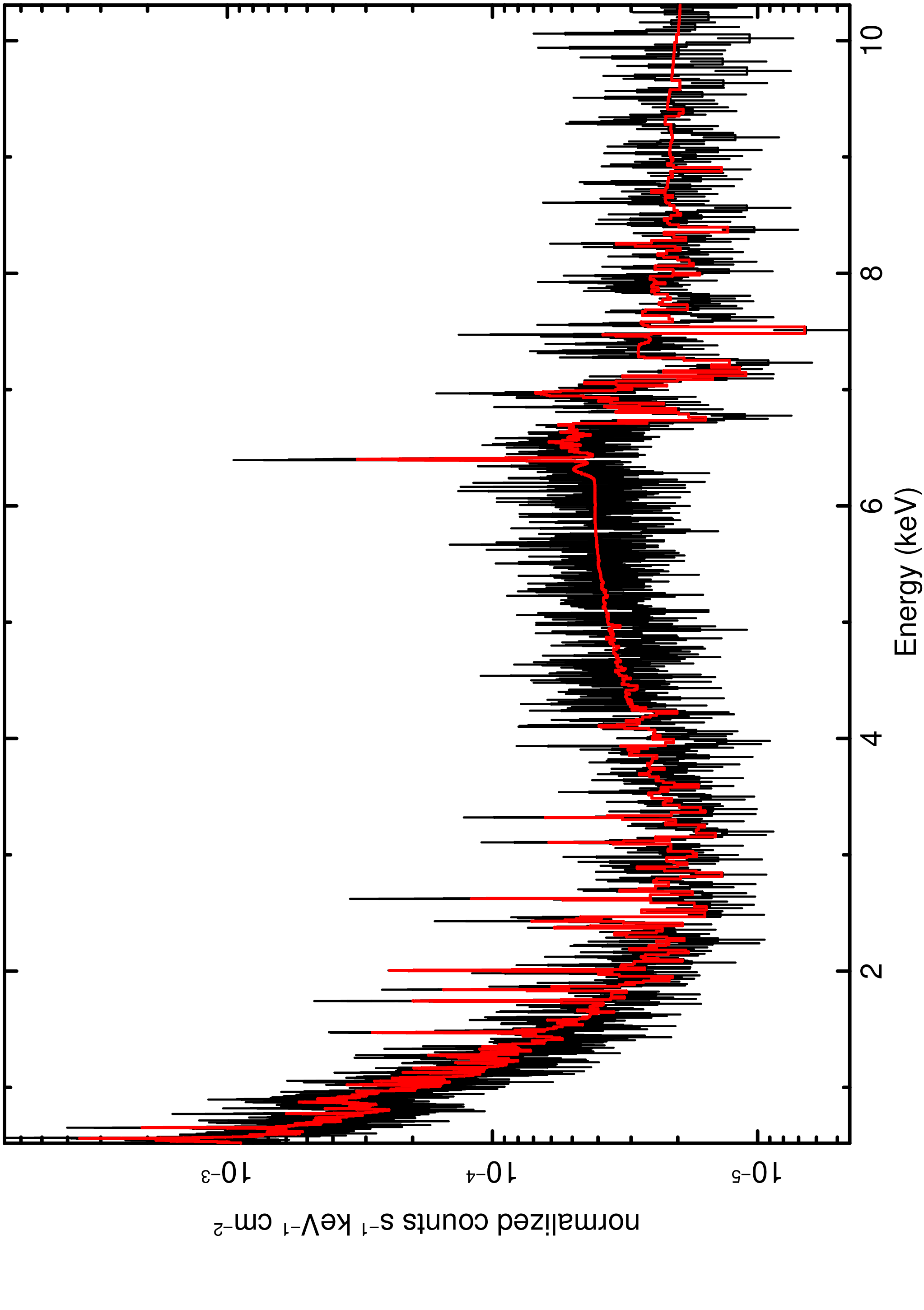}
    \caption{These panels show simulated 100~ks XRISM spectra of Mrk~817, made assuming the best-fit model detailed in Table 1.  At CCD resolution, the fastest outflows in Mrk 817 are apparent, but the anticipated 5~eV resolution of the Resolve calorimeter spectrometer makes the strong, highly blue-shifted absorption lines in the Fe~K band especially clear.  These panels can be directly compared with those in Figure 4.}
    \label{fig:xrism}
\end{figure*}

\end{document}